\documentclass[final,3p,times,pdflatex]{elsarticle}
\usepackage{amsmath}
\usepackage{amssymb}
\usepackage{graphicx}
\usepackage{color} 
\definecolor{Red}{cmyk}{0,1,1,0}
\definecolor{Blue}{cmyk}{1,0.75,0,0}

\newcommand\beq{\begin{eqnarray}}
\newcommand\eeq{\end{eqnarray}}

\def\SOFTSUSY{{\tt SOFTSUSY}}
\def\code#1{{\tt #1}}

\journal{Computer Physics Communications}

\begin{document}

\begin{frontmatter}

\begin{flushright}
DAMTP-2016-16\\
IFIC/??
\end{flushright}

\title{The Inclusion of Two-Loop SUSYQCD Corrections to Gluino and Squark 
    Pole Masses in the Minimal and Next-to-Minimal Supersymmetric Standard Model: {\tt       SOFTSUSY3.7}}

\author[damtp]{B.C.~Allanach}
\cortext[cor1]{Corresponding author}
\author[illinois]{Stephen~P.~Martin}
\author[otterbein]{David~G.~Robertson}
\author[valencia]{Roberto~Ruiz~de~Austri\corref{cor1}}
\ead{rruiz@ific.uv.es}

\address[damtp]{DAMTP, CMS, University of Cambridge, Wilberforce road,
  Cambridge, CB3  0WA, United Kingdom}
\address[illinois]{Department of Physics, Northern Illinois University, DeKalb, IL 60115, United States of America
}
\address[otterbein]{Department of Physics, Otterbein University, Westerville,
  OH 43081, United States of America}
  \address[valencia]{Instituto de F\'i�sica Corpuscular, IFIC-UV/CSIC, E-46980
  Paterna, Spain}  
\begin{abstract}
We describe an extension of the {\tt SOFTSUSY} spectrum calculator to include
two-loop 
supersymmetric QCD (SUSYQCD) 
corrections 
of order $\mathcal{O}(\alpha_s^2)$
to gluino and squark pole
masses, either in the minimal supersymmetric standard model (MSSM) or the
next-to-minimal supersymmetric standard model (NMSSM). This document
provides an overview of the program and acts as a manual for the
new version of {\tt SOFTSUSY}, which includes the increase in accuracy in
squark and gluino pole mass predictions. 
\end{abstract}

\end{frontmatter}

\section{Program Summary}
\noindent{\em Program title:} \SOFTSUSY{} \\
{\em Program obtainable   from:} {\tt http://softsusy.hepforge.org/} \\
{\em Distribution format:}\/ tar.gz \\
{\em Programming language:} {\tt C++}, {\tt fortran}, {\tt C} \\
{\em Computer:}\/ Personal computer. \\
{\em Operating system:}\/ Tested on Linux 3.4.6, 
Mac OS X 10.7.5
\\
{\em Word size:}\/ 64 bits. \\
{\em External routines:}\/ None \\
{\em Typical running time:}\/ 15 seconds per parameter point. \\
{\em Nature of problem:}\/ Calculating supersymmetric particle spectrum, 
mixing parameters and couplings in the MSSM or the NMSSM\@. The solution to
the renormalisation group equations must be consistent 
with theoretical boundary conditions on supersymmetry breaking parameters, as
well as a weak-scale boundary condition on gauge 
couplings, Yukawa couplings and the Higgs potential parameters. \\
{\em Solution method:}\/ Nested fixed point iteration.  \\
{\em Restrictions:} \SOFTSUSY~will provide a solution only in the
perturbative regime and it
assumes that all couplings of the model are real
(i.e.\ $CP-$conserving). If the parameter point under investigation is
non-physical for some reason (for example because the electroweak potential
does not have an acceptable minimum), \SOFTSUSY{} returns an error message.
The higher order corrections included are for the 
MSSM ($R-$parity conserving or violating) or the real $R-$parity conserving
NMSSM only. \\
{\em CPC Classification:}\/ 11.1 and 11.6. \\
{\em Does the new version supersede the previous version?:}\/ Yes. \\
{\em Reasons for the new version:}\/ It is desirable to improve the accuracy of
the squark and gluinos mass predictions, since they strongly affect
supersymmetric particle production cross-sections at colliders.  \\
{\em Summary of revisions:}\/
The calculation of the squark and gluino pole masses is extended to be of
next-to-next-to leading order in SUSYQCD, i.e.\ including terms up to ${\mathcal
  O}(g_s^4/(16 \pi^2)^2)$.

\section{Introduction}

Near the beginning of LHC Run II at 13 TeV centre of mass collision energy,
hopes are high for the discovery of new physics. A much-studied and
long-awaited framework for new physics, namely weak-scale superymmetry, is being
searched for in many different channels. In the several prominent explicit
models of 
supersymmetry breaking meditaion, the best chance of observing the production
and subsequent decay of supersymmetric particles is via squark and/or gluino
production. Squarks and gluinos tend to have the largest
production cross-sections among the sparticles and Higgs bosons of the MSSM,
because they may be produced by tree-level strong interactions, as opposed to
smaller electroweak cross sections relevant for the other superparticles. 
In $R-$parity conserving supersymmetry (popular because of its
apparently viable dark matter candidate), squark and/or gluino production
may result in a signal of an excess of highly energetic jets in conjunction
with large missing 
transverse momentum with respect to Standard Model predictions. This classic
LHC signature was searched for at LHC Run I 
at 7 and 8 TeV centre of mass energy, but no significant excess above Standard
Model backgrounds was observed. 
Jets plus missing transverse momentum channels then provided the strongest
constraints 
in the constrained minimal supersymmetric standard model, for instance
(CMSSM), along with many other models of supersymmetry breaking mediation.

The higher collision energy at Run II allows for new MSSM parameter space to
be explored. In the event of a discovery of the production of supersymmetric
particles, one will want to first interpret their signals correctly, and then
make inferences about the superymmetry breaking parameters. In order to do
this with a greater accuracy, we must use higher orders in perturbation
theory. Of particular interest is the connection between the supersymmetric
masses of the squarks and gluinos and the experimental observables (functions
of jet and missing momenta). The experimental observables may be used to infer
{\em pole}\/ (or kinematic) masses of the supersymmetric particles, which may
then be connected via perturbation theory to the more fundamental
supersymmetry breaking parameters in the Lagrangian. These could then be used
to test the underlying supersymmetry breaking mediation
mechanism~\cite{Allanach:2004ud,Allanach:2004ed}.
Conversely, if no significant signal for supersymmetry is to be found at the
LHC, we shall want to interpret the excluded parameter space in terms of the
fundamental supersymmetry breaking parameters. Again, this connection is
sensitive (for identical reasons to the discovery case) to the order in
perturbation theory which is used. In the most easily accessible channels at
the LHC (squark/gluino production), it is useful therefore to use higher
orders in the QCD gauge coupling, since this is the largest relevant expansion
parameter. 

State-of-the art publicly available NMSSM or MSSM spectrum generators such as
{\tt ISAJET}~\cite{Paige:2003mg}, {\tt FlexibleSUSY}~\cite{Athron:2014yba},
{\tt NMSPEC}~\cite{Ellwanger:2006rn}, {\tt 
  SUSPECT}~\cite{Djouadi:2002ze}, {\tt SARAH}~\cite{Staub:2008uz}, 
{\tt SPHENO}~\cite{Porod:2003um}, {\tt SUSEFLAV}~\cite{Chowdhury:2011zr} or
previous versions of {\tt 
  SOFTSUSY}~\cite{Allanach:2001kg}, 
 do not have
the complete 
${\mathcal{O}}(\alpha_s^2/(16 \pi^2))$ two-loop corrections to gluino and
squark masses included, despite their being calculated and presented in the
literature~\cite{Martin:2005ch,Martin:2005eg,Martin:2006ub}. Here, we describe
their inclusion into the MSSM  
spectrum calculation in the popular {\tt SOFTSUSY} program, making them
publicly available for the first time. As we emphasised above, we expect them
to be useful in increasing the accuracy of inference from data of
supersymmetry breaking in the squark and gluino sectors. 

The paper proceeds as follows: in the next section, the higher order terms
that are included are briefly reviewed. We
then provide an example of their effect on a line through CMSSM
space. After a summary, the appendices contain technical information on how to
compile and run {\tt SOFTSUSY} including the higher order terms. 

\section{Higher Order Terms  \label{sec:results}} 

Two-loop contributions to fermion pole masses in gauge theories were
calculated in Ref.~\cite{Martin:2005ch} and these results are specialised to
compute the gluino pole mass. Depending upon the ratio $m_{\tilde
  q}/m_{\tilde g}$, a two-loop correction of up to several percent was
found.
On the other hand, Ref.~\cite{Martin:2005eg} calculated the two-loop
contributions to scalar masses from gauge theory
and these results are
specialised to the two-loop squark masses, where corrections up to
about one percent were noted. 
In both cases, the results were obtained in the 
$\overline{\rm DR}'$ renormalization scheme \cite{Jack:1994rk}, consistent with the 
renormalization group equations used in softly broken supersymmetric models.
The version of {\tt SOFTSUSY} described here now contains these
computations~\cite{Martin:2005ch,Martin:2005eg}. 
A library for computing two-loop self-energy integrals, {\tt
TSIL}~\cite{Martin:2005qm}, is included within the {\tt SOFTSUSY}
distribution; there is no need to download it separately.

In addition, the gluino result has been improved by re-expanding the gluino 
self-energy function squared mass arguments around the gluino, squark, and top quark pole masses, as described in Ref.~\cite{Martin:2006ub}. This requires iteration to 
determine the gluino pole mass, and hence is slower than simply evaluating the self-energy functions with 
the lagrangian mass parameters, but the accuracy of the calculation is improved 
significantly~\cite{Martin:2006ub}.  
Typically, 4 to 6 iterations are required to reach the 
default tolerance of less than 1 part in $10^5$ difference between
successive iterations for the gluino pole mass.
This iteration is not the bottleneck in the calculation, as we find that the CPU time
spent on the squark masses is about twice that spent on the gluino mass, even with 6 iterations for the latter. 
We did not implement this improvement for the squarks since the effect is less 
significant, and the necessary iterations would be much slower.

\subsection{Kinky Masses\label{subsec:kinky}}
\begin{figure}[!tb]
\centering
\includegraphics[width=12.0cm,angle=0]{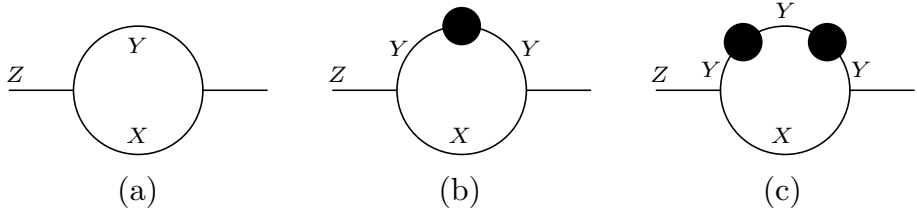}
\caption{\label{fig:kinks}
Feynman diagrams for the self-energy of a particle that couples to
two lighter particles. The dark blobs in (b) and (c) represent 1-loop subdiagrams.
Near threshold, diagrams (b) and (c) have $1/(-\delta)^{1/2}$ and $1/(-\delta)^{3/2}$
singularities, respectively, where $s = (\sqrt{x}+\sqrt{y})^2(1 + \delta)$, with
$x$ and $y$ the squared masses of the internal particles.
}
\end{figure}

In implementing these 2-loop pole masses, we encountered an interesting issue that
does not seem to have been noted before, as far as we know. Consider the self-energy
and pole mass of a particle $Z$ that has a three-point coupling to particles $X$ and $Y$.
Let the tree-level squared masses of these particles by $z$, $x$, and $y$ respectively.
If $z$ happens to be close to the threshold value 
$(\sqrt{x}+\sqrt{y})^2$, then the computed 2-loop complex pole mass of $Z$ will have 
a singularity proportional to
\beq
1/\sqrt{-\delta - i \epsilon}
\eeq
where $\epsilon$ is infinitesimal and positive, and
\beq
z = (\sqrt{x}+\sqrt{y})^2 (1 + \delta).
\eeq
The reason for this can be understood from the sequence of
Feynman diagrams shown in Figure~\ref{fig:kinks}. After reduction to basis integrals,
the result of Figure~\ref{fig:kinks}(b) contains terms proportional to $V(x,y,u,v)$ and
$B(x,y')$, in the notation of refs.~\cite{Martin:2003qz,Martin:2005qm}. (Here the prime represents a derivative with respect 
to the corresponding argument, and the external momentum invariant is $s=z$.)
Then, for example, one can evaluate:
\beq
B(x,y') \Big |_{s = (\sqrt{x} + \sqrt{y})^2(1 + \delta)} &=&
-\frac{\pi (x/y)^{1/4}}{(\sqrt{x}+\sqrt{y})^2 \sqrt{-\delta - i \epsilon}}
+ \ldots~.
\eeq
We have found that in general such singularities do not cancel within the fixed-order
2-loop pole mass
calculation. This includes e.g.\ the pole mass of the gluino, where we checked that there are singularities in the pole mass as one varies the top mass and one of the stop masses very close to the 2-body decay threshold. 
Similarly, there are singularities in the 2-loop pole masses of the top squarks, 
in each case when the top mass and the gluino happen to be very close to the 2-body decay thresholds. 
We have also checked that this behavior occurs
in a simple toy model involving only three massive scalar particles. 

This singular behavior might be quite surprising, because the pole mass is
supposed to be an observable, and therefore ought to be free of divergences of any kind.
The resolution is that, similar to problems with infrared divergences, the singularity
is an artifact of truncating perturbation theory. The 3-loop diagram of
Figure~\ref{fig:kinks}(c) will diverge  
like $1/(-\delta)^{3/2}$, and similar diagrams of loop order 
$L$ will diverge like $1/(-\delta)^{L-3/2}$. 
These contributions can presumably be resummed to 
give a result that is well-behaved as $\delta \rightarrow 0$, although proving
that is beyond the 
scope of the present paper.

\begin{figure}[!tb]
\centering
\includegraphics[width=4.0cm,angle=0]{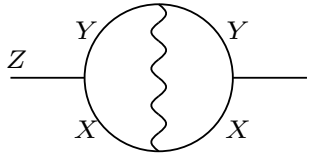}
\caption{\label{fig:kinks2}
Feynman diagram for the self-energy of a particle that couples to
two lighter particles, each of which couples to a massless gauge boson. 
Near threshold, this diagram has a $\ln(-\delta)$ singularity, where 
$s = (\sqrt{x}+\sqrt{y})^2(1 + \delta)$, with
$x$ and $y$ the squared masses of the internal particles.}
\end{figure}

The above singularity behavior is not tied to the presence 
of massless gauge bosons. However, if massless gauge bosons are present, 
then there is another, less severe, type of singular threshold behavior, due to 
the Feynman diagram shown in Figure~\ref{fig:kinks2}.
In the notation of refs.~\cite{Martin:2003qz,Martin:2005qm}, 
this arises due to the near-threshold dependence of the corresponding 
self-energy basis integral:
\beq
M(x,x,y,y,0) \Big |_{s = (\sqrt{x} + \sqrt{y})^2(1 + \delta)} &=&
-\frac{2\pi^2}{(\sqrt{x}+\sqrt{y})^2} \ln(-\delta - i \epsilon)
+ \ldots~.
\eeq 
Again, the singularity in the computed 2-loop pole mass
is an artifact of the truncation of perturbation theory, and would presumably be removed
by a resummation including all diagrams with $2,3,4,\ldots$ massless gauge bosons exchanged between particles $X$ and $Y$.

Note that it requires some bad luck to encounter any of these threshold 
singularity problems in 
practice, as there is
no good reason why the tree-level masses should be tuned to the high 
precision necessary to make the problem numerically significant. Nevertheless, it could
lead to a small\footnote{Note that 
(before smoothing by the interpolation method adopted here) 
this kink in the 2-loop computed pole mass is 
actually infinitely large as $\delta\rightarrow 0$, but in practice it is confined to a 
small region, because of the 2-loop suppression factor. Therefore, it would usually appear to be finite in any scan with reasonable increments in masses.} 
``kink" in the computed pole mass if one performs a scan over models by varying the 
input masses. To avoid this possibility, in our code we test for small $\delta$, and 
then replace the offending 2-loop self-energy basis integrals $B(x,y')$
and $V(x,y,u,v)$ 
and $M(x,x,y,y,0)$
by values interpolated from slightly
larger and smaller values of the external momentum invariant. 
Explicitly, if $\delta<t$ (we choose 
$t=0.04$
in the current version 
of SOFTSUSY, with the value controlled by the quantity \code{THRESH\_TOL} in
\code{src/supermodel/supermodel/sumo\_params.h}), 
the integral in question $I(s)$ is 
replaced by 
\begin{equation}
I(s) \approx \frac{1}{2} \left(1 + \frac{\delta}{t}\right) I\left( [1+t]s\right) + 
\frac{1}{2} \left(1 - \frac{\delta}{t}\right) I\left( [1-t]s\right).
\end{equation}
This provides a
pole mass 
that varies continuously  in scans of input masses near thresholds, and the
difference between our value and the value that would be obtained by a proper
resummation should be small.

\begin{figure}[!t]
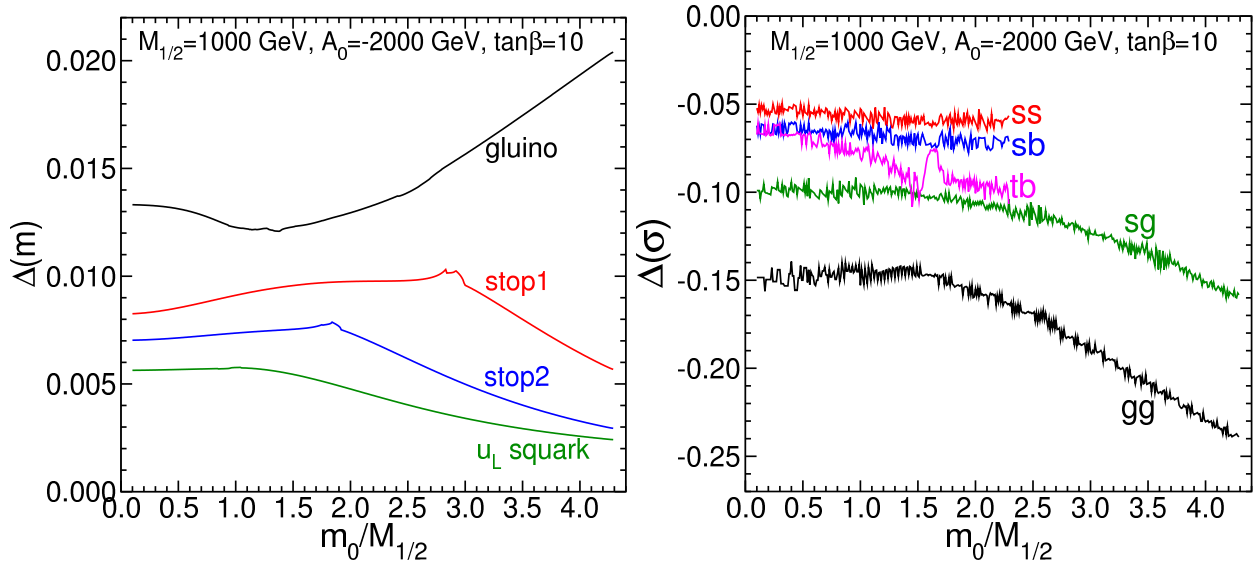

\includegraphics[width=0.495\textwidth]{deltam.eps}
\includegraphics[width=0.495\textwidth]{deltasigma.eps}
\caption{\label{fig:m0scan} 
Effects of the 2-loop SUSYQCD corrections to sparticle pole
  masses (left panel) and total production cross-sections (right panel), for a
  line in CMSSM space with $M_{1/2}=1$ TeV, $A_0=-2$ TeV, $\tan \beta=10$ and
  $\mu>0$.  In the left panel, $\Delta(m)=m_\textrm{2-loop} / m_\textrm{1-loop} - 1$, where
$m_\textrm{2-loop}$ is the pole mass calculated including the
${\mathcal{O}}(\alpha_s^2/(16 \pi^2))$ terms whereas 
$m_\textrm{1-loop}$ is the pole mass calculated at one-loop order. 
In the right panel, $\Delta(\sigma)=\sigma_\textrm{2-loop} / \sigma_\textrm{1-loop} - 1$, where
$\sigma_\textrm{2-loop}$ is the 13 TeV LHC production cross-section calculated 
using the ${\mathcal{O}}(\alpha_s^2/(16 \pi^2))$ terms in pole masses whereas 
$\sigma^\textrm{1 loop}$ is the 13 TeV LHC production cross-section calculated
with one-loop pole masses. The label `gg' refers to gluino-gluino production, `sg' to
squark-gluino plus antisquark-gluino, `ss' to squark-squark plus antisquark-antisquark, `sb' to squark-antisquark, and
`tb' to stop-antistop production. The cross-sections in each case are obtained
at NLL+NLO
using {\tt NLL-fastv3.1-13TeV}
\cite{Beenakker:2015rna,Beenakker:2011fu,Beenakker:2010nq,Beenakker:1997ut,Beenakker:2011fu,Beenakker:2009ha,Kulesza:2009kq,Kulesza:2008jb,Beenakker:1996ch}. 
\label{fig:reldiff}.}
\end{figure}
\subsection{Illustration of Results}
We now illustrate the effect of the ${\mathcal{O}}(\alpha_s^2/(16 \pi^2))$
corrections to gluino and squark pole masses, taking the constrained minimal
supersymmetric standard model (CMSSM) pattern of MSSM supersymmetry breaking
Lagrangian terms. In the CMSSM, at the gauge unification scale 
$\sim {\mathcal O}(10^{16}) \textrm{~GeV}$ the gaugino masses are set equal to $M_{1/2}$,
the scalar masses are set to a universal flavour diagonal mass $m_0$ and the
soft supersymmetry breaking trilinear scalar couplings are set equal to a
massive parameter $A_0$ times the relevant Yukawa coupling. Here, for
illustration, we set $M_{1/2}=1$ TeV, $A_0=
-2\>{\rm TeV}$, the ratio of the MSSM's two
Higgs vacuum expectation values $\tan \beta=10$ and the sign of the Higgs
superpotential parameter term $\mu$ to be positive. We then allow $m_0$ to
vary, and plot the relative difference caused by the new higher order terms in
 Fig.~\ref{fig:reldiff}a. 
For this illustration, we have {\em not}\/
included two-loop corrections to gauge and
Yukawa couplings and three-loop renormalisation group equations for the
superpotential parameters~\cite{Allanach:2014nba}, although with them, the
results are qualitatively similar.
The overall message from the figure is that differences of percent level order
in the
pole masses of gluinos and squarks arise from the higher order corrections. 
In Fig.~\ref{fig:reldiff}a, the plot does not extend to larger values of
$m_0/M_{1/2}$, because there is no 
phenomenologically acceptable electroweak symmetry breaking there (the
superpotential $\mu$ term becomes imaginary at the minimum of the potential,
indicating a saddle point). 
The rough size of the two-loop correction is consistent with that
estimated in the
previous literature~\cite{Martin:2005ch,Martin:2005eg,Martin:2006ub}.
Note that there are wiggles in the $\tilde g$, $\tilde t_1$, and $\tilde t_2$ curves,
near $m_0/M_{1/2} = 1.35$, $2.9$, and $1.9$, respectively. These are the remnants of
the instabilities mentioned in section~\ref{subsec:kinky} after smoothing by the interpolation procedure described there, with $t=0.04$. 
These correspond to the thresholds for the
decays $\tilde g \rightarrow t \tilde t_1$ and
$\tilde t_1 \rightarrow t \tilde g$ and $\tilde t_2 \rightarrow t \tilde g$, respectively. 
Although the remaining wiggles are visible on the plot, they are 
small in absolute terms, and are of order the uncertainties 
due to higher-order corrections.

In Fig.~\ref{fig:reldiff}b, we show the concomitant effect on the various
next-to-leading order
sparticle production cross-sections at the 13 TeV LHC by 
using {\tt
  NLL-Fast3.1}~\cite{Beenakker:2015rna,Beenakker:2011fu,Beenakker:2010nq,Beenakker:1997ut,Beenakker:2011fu,Beenakker:2009ha,Kulesza:2009kq,Kulesza:2008jb,Beenakker:1996ch}. {\tt
  NLL-Fast3.1} calculates at the next-to-leading order (NLO) in supersymmetric QCD,
with next-to-leading logarithm (NLL) re-summation. However, 
it is based on fixed interpolation tables with only 3 significant digits,
resulting in a visible jaggedness of the points in the figure. 
Nonetheless, we see that the change in the gluino mass due to 
next-to-next-to leading order (NNLO) effects
leads to a large 
15-25$\%$ 
reduction in the production cross-section. 
Other sparticle production cross-section modes shown decrease by more than
5$\%$, thus accounting for these NNLO effects is important in reducing the
theoretical uncertainties. Some of the curves terminate when $m_0/M_{1/2}>2.3$
because {\tt NLL-Fast3.1} considers the production cross-sections to be too
small to be relevant, and so returns zeroes. 

\begin{figure}[!t]
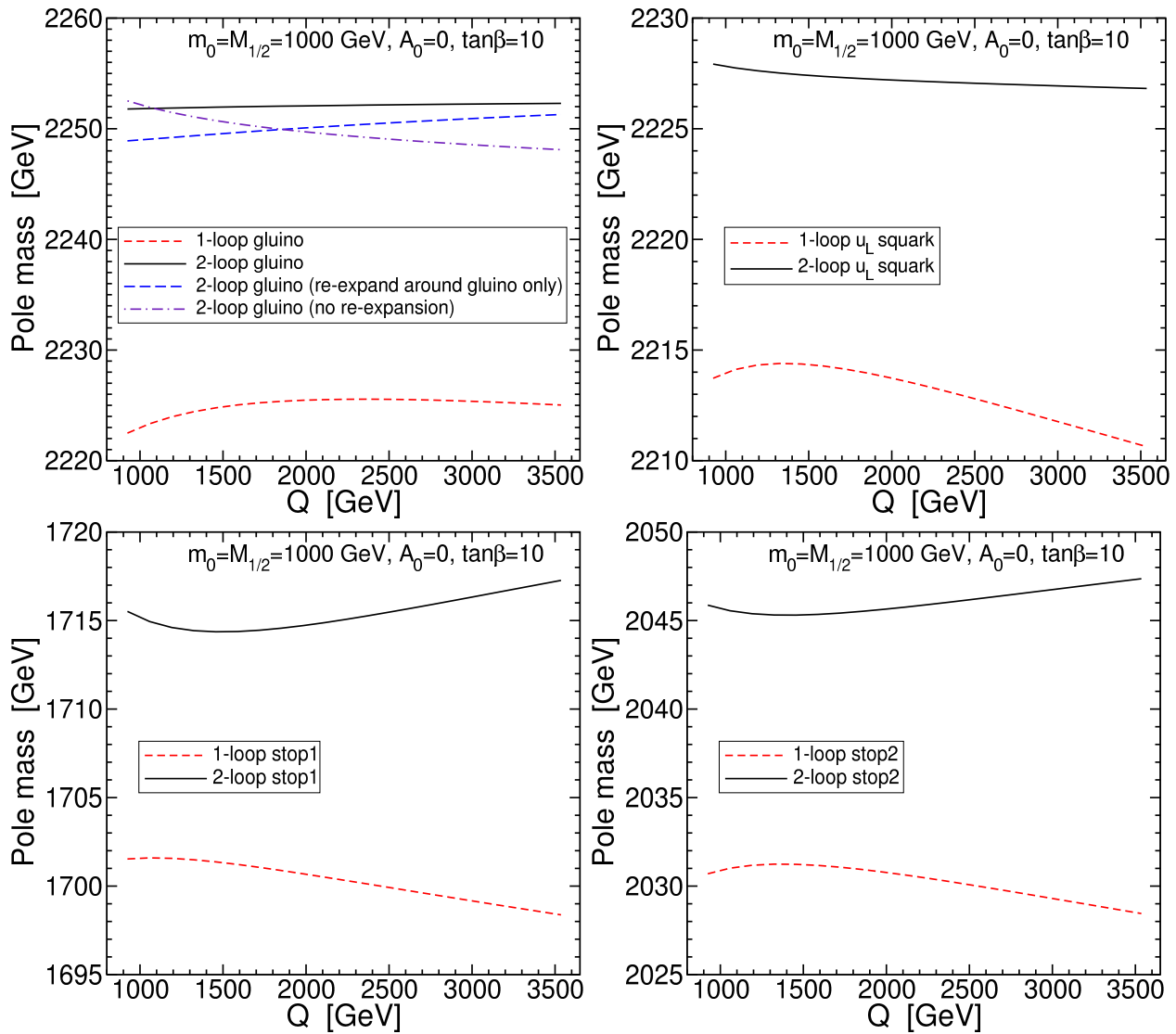

\includegraphics[width=0.495\textwidth]{Qgluino.eps}
\includegraphics[width=0.495\textwidth]{QsuL.eps}
\includegraphics[width=0.495\textwidth]{Qstop1.eps}
\includegraphics[width=0.495\textwidth]{Qstop2.eps}
\caption{\label{fig:Qdep} 
Scale dependences of gluino and squark pole mass
predictions for the CMSSM with $m_0=M_{1/2}=1$ TeV, $A_0=0$, $\tan \beta=10$
and $\mu>0$, comparing 1-loop and 2-loop approximations. The upper left panel is
for the gluino, upper right panel for $\tilde u_L$, lower left panel for $\tilde t_1$,
and lower right panel for $\tilde t_2$. In the case of the gluino, the solid line 
uses the re-expansion of squared mass arguments about the gluino and squark pole 
masses as described in \cite{Martin:2006ub};
this is the default used by SOFTSUSY for the gluino 
when the higher-order pole mass calculations are enabled.
}
\end{figure}
In Fig.~\ref{fig:Qdep}, we illustrate the renormalization scale $(Q)$ dependences 
of the computed gluino and squark pole masses, 
by varying the scale $Q$ at which the masses are calculated
by a factor of 2 around $M_{SUSY}=\sqrt{m_{{\tilde t}_1} m_{{\tilde t}_2}}$. 
We see the expected reduced scale dependence going from 1-loop to 2-loop order. 
In the case of the gluino,  we also see that the 2-loop calculation
using a re-expansion around both the squark and gluino pole masses,
as described in \cite{Martin:2006ub},
displays less renormalization scale dependence than the case where only the
gluino mass is re-expanded around its pole mass in the gluino pole mass calculation, or
the case in which no re-expansion is used.   
This is consistent with the
suggestion in \cite{Martin:2006ub} that the perturbation expansion is made
more convergent by re-expanding around pole masses rather than around the running masses. 
For this reason, the re-expansion method is used as the default by SOFTSUSY when the
higher-order pole masses are enabled.
In the cases of the squark masses, the improvement of the renormalization scale dependence 
of the computed pole mass is less significant going from 1-loop to 2-loop order. 
We also note that in each case, the 2-loop correction is much larger than the 
scale dependence in the 1-loop result. Thus, as usual, the renormalization scale 
dependence gives only a lower bound, and not a reliable estimate, of the remaining 
theoretical error.

\section{Summary and Conclusions}
Two-loop ${\mathcal{O}}(\alpha_s^2/(16 \pi^2))$ are now included in the public
release of {\tt \small SOFTSUSY}, and so are available for use. We have
demonstrated that along a typical line in CMSSM space, they are responsible
for around $\mathcal{O}(1\%)$ relative changes in the squark and gluino
masses, which will change their various production cross-sections by around
$5-25\%$. 
The largest 2-loop SUSYQCD pole mass correction is for the gluino,
and it increases as the squarks become heavier than the gluino, reaching 2\%
when $m_0/M_{1/2} = 4$.
These sources of theoretical error can now be taken into account and 
consequently reduced by using the new version of {\tt SOFTSUSY}. In turn, the
connection between measurements and various Lagrangian parameters (in
particular, soft supersymmetry breaking sparticle mass parameters) is made
more accurate. Thus, fits of the MSSM or NMSSM to cosmological and collider
data will have a smaller associated theoretical error (as would studies of the
unification of sparticle masses were there to be a discovery and subsequent
measurement of sparticles).
Other increases in MSSM or NMSSM mass prediction accuracy await future work:
Higgs mass predictions, gauginos and sleptons, for instance. 

\section*{Acknowledgments}
This work has been partially supported by STFC grant 
ST/L000385/1. We thank the Cambridge SUSY working group for helpful
discussions. 
The work of SPM was supported in part by the National Science
Foundation grant number PHY-1417028.
DGR acknowledges the support of the Ohio Supercomputer Center. 
The work of R. RdA is supported by the Ram\'on y Cajal program of the Spanish MICINN 
and also thanks the support of the Spanish MICINN's Consolider-Ingenio 2010 Programme 
under the grant MULTIDARK CSD2209-00064, the Invisibles European 
ITN project (FP7-PEOPLE-2011-ITN, PITN-GA-2011-289442-INVISIBLES and the 
``SOM Sabor y origen de la Materia" (FPA2011-29678) and the 
``Fenomenologia y Cosmologia de la Fisica mas alla del Modelo Estandar e lmplicaciones Experimentales 
en la era del LHC" (FPA2010-17747) MEC projects. 

\appendix

\section{Installation of the Increased Accuracy Mode}
\label{sec:install}

In order to have {\tt \small SOFTSUSY} use the higher order corrections
described in the present paper, the code containing them must be both compiled
{\em and}\/ a run-time flag should be set to ensure their employment in the
spectrum calculation. 
A compilation argument to the {\tt ./configure}~command is provided in order
to compile the necessary code to include the higher order corrections:
\begin{verbatim}
--enable-two-loop-sparticle-mass-compilation
\end{verbatim}
We have included a global boolean variable which controls the higher order
corrections  at run time (provided that the program has already been compiled
with  the higher order corrections included):
\begin{quote}
 \verb|bool USE_TWO_LOOP_SPARTICLE_MASS|  - if \code{true}, the 
          $\mathcal{O}(\alpha_s^2)$ corrections are included 
          (corresponds to the \code{SOFTSUSY Block} parameter \code{22} in the
          {\tt SOFTSUSY} block of the SUSY Les Houches Accord input). 
\end{quote}
By default, the higher order corrections are switched off
(the boolean value is set to {\tt false}), unless the user sets it in
their main program, or in the input parameters (see~\ref{sec:run}). 
One can choose to include the two-loop $\mathcal{O}(\alpha_s^2)$ corrections
to gluino and squark pole masses independently of whether one includes
two-loop corrections to the extracted MSSM value of $\alpha_s(M_Z)$ or
three-loop MSSM renormalisation group equations, as
described in Ref.~\cite{Allanach:2014nba}.
        
To summarise, installation is completed by executing the following commands
\begin{verbatim}
> ./configure --enable-two-loop-sparticle-mass-compilation
> make
\end{verbatim}
We remind the reader that the two-loop corrections discussed here are
available for use either in the MSSM (with or without $R-$parity
violation)~\cite{Allanach:2009bv} or 
in the NMSSM~\cite{Allanach:2013kza}.

\section{Running \SOFTSUSY~in the Increased Accuracy Mode}  
\label{sec:run}

\SOFTSUSY~produces an executable called \code{softpoint.x}. 
One can run this executable from command line arguments, but the higher order
corrections will be, by default, switched off. One may switch the two-loop
$\mathcal{O}(\alpha_s^2)$ corrections described in the present paper on with
the argument 
\verb|--two-loop-sparticle-masses|.

For example: 
{\small\begin{verbatim}
./softpoint.x sugra --tol=1.0e-5 --m0=250 --m12=100 --a0=-100 --tanBeta=10 --sgnMu=1 \
--two-loop-sparticle-masses --two-loop-sparticle-mass-method=<expandAroundGluinoPole>
\end{verbatim}\normalsize}
The variable \code{expandAroundGluinoPole}
re-expands the two-loop computation of the gluino mass around the
gluino and squark masses if it is set to \code{3} (default), around only the
gluino mass 
if set to \code{2} and performs no expansion (but still includes the 2-loop
corrections) if it is set to \code{1}.

For the calculation
of the spectrum of single points in parameter space, one could alternatively
use the 
SUSY Les Houches Accord (SLHA)~\cite{Skands:2003cj,Allanach:2008qq}
input/output 
option. The user must provide a file (e.g.\ the example file included
in the \SOFTSUSY~distribution
\code{inOutFiles/lesHouchesInput}) which specifies the model dependent input
parameters. The program may then be run with
\small
\begin{verbatim}
 ./softpoint.x leshouches < inOutFiles/lesHouchesInput
\end{verbatim}
\normalsize

One can change whether the higher order corrections are switched on 
(provided they have been compiled by
setting the correct \verb|./configure| flag as described above)
with
\code{SOFTSUSY Block} parameter 22 and the two-loop gluino expansion
approximation with parameter 23:
\begin{verbatim}
Block SOFTSUSY               # Optional SOFTSUSY-specific parameters
   22   1.000000000e+00      # Include 2-loop terms in gluino/squark masses 
                             # (default of 0 to disable)
   23   3.000000000e+00      # sets expandAroundGluinoPole parameter (default 3)
\end{verbatim}
Parameter 23 is equal to the integer global variable
\code{expandAroundGluinoPole}.

We are also providing an example user program called \code{higher.cpp}, 
found in the \code{src/} directory of the SOFTSUSY distribution. Running 
\code{make} in the main SOFTSUSY directory produces an executable called \code{higher.x}, 
which runs without arguments or flags. 
This program illustrates the implementation of the 2-loop 
SUSYQCD pole masses, and in particular outputs all of the 
data used in Figures~\ref{fig:m0scan} and~\ref{fig:Qdep} above. The 
file \code{higher.cpp} and the output of \code{higher.x} (called \code{twoLoop.dat})
are also found as ancillary electronic files with the {\tt arXiv} submission for this article.

\bibliography{ho}
\bibliographystyle{elsarticle-num}
\end{document}